\documentclass[aps,onecolumn]{revtex4-2}
\usepackage{graphicx}
\usepackage[english]{babel}
\usepackage[utf8x]{inputenc}
\usepackage[T1]{fontenc}
\usepackage{mathtools}
\usepackage{amsmath}

\usepackage{physics}
\usepackage{ upgreek }
\usepackage{wrapfig}
\usepackage{color}
\usepackage{soul}
\usepackage[letterpaper]{geometry}
\usepackage{bm}
\usepackage{caption}
\usepackage{float}
\usepackage{stfloats}
\usepackage{listings}

\DeclareFixedFont{\ttb}{T1}{txtt}{bx}{n}{12} 
\DeclareFixedFont{\ttm}{T1}{txtt}{m}{n}{12}  

\usepackage{color}
\definecolor{deepblue}{rgb}{0,0,0.5}
\definecolor{deepred}{rgb}{0.6,0,0}
\definecolor{deepgreen}{rgb}{0,0.5,0}
\definecolor{blue}{rgb}{0,0,1}

\newcommand\pythonstyle{\lstset{
language=Python,
basicstyle=\small,
otherkeywords={self},             
keywordstyle=\bfseries\color{deepblue},
emph={Ball,__init__},          
emphstyle=\bfseries\color{deepred},    
stringstyle=\color{deepgreen},
frame=tb,                         
showstringspaces=false,            %
commentstyle=\color{blue}
}}

\lstnewenvironment{python}[1][]
{
\pythonstyle
\lstset{#1}
}
{}

\newcommand\pythoninline[1]{{\pythonstyle\lstinline!#1!}}

\newsavebox\IBoxA \newsavebox\IBoxB \newlength\IHeight
\newcommand\TwoFig[6]{
  \sbox\IBoxA{\includegraphics[width=0.45\textwidth]{#1}}
  \sbox\IBoxB{\includegraphics[width=0.45\textwidth]{#4}}%
  \ifdim\ht\IBoxA>\ht\IBoxB
    \setlength\IHeight{\ht\IBoxB}%
  \else\setlength\IHeight{\ht\IBoxA}\fi
  \begin{figure}[!htb]
  \minipage[t]{0.45\textwidth}\centering
  \includegraphics[height=\IHeight]{#1}
  \caption{#2}\label{#3}
  \endminipage\hfill
  \minipage[t]{0.45\textwidth}\centering
  \includegraphics[height=\IHeight]{#4}
  \caption{#5}\label{#6}
  \endminipage 
  \end{figure}%
}
\usepackage{subcaption}
\usepackage{amsmath}
\usepackage{graphicx}

\begin{document}

\title{Jamming and arrest of cell motion in biological tissues}
\author{Elizabeth Lawson-Keister and M. Lisa Manning}
\email{Corresponding author: Manning, M. Lisa (mmanning@syr.edu)}
\affiliation{Department of Physics and BioInspired Institute, Syracuse University, Syracuse, NY 13244, USA}
\maketitle
    Collective cell motility is crucial to many biological processes including morphogenesis, wound healing, and cancer invasion. Recently, the biology and biophysics communities have begun to use the term “cell jamming” to describe the collective arrest of cell motion in tissues. Although this term is widely used, the underlying mechanisms are varied. In this review, we highlight three independent mechanisms that can potentially drive arrest of cell motion -- crowding, tension-driven rigidity, and reduction of fluctuations -- and propose a speculative phase diagram that includes all three. Since multiple mechanisms may be operating simultaneously, this emphasizes that experiments should strive to identify which mechanism dominates in a given situation.  We also discuss how specific cell-scale and molecular-scale biological processes, such as cell-cell and cell-substrate interactions, control aspects of these underlying physical mechanisms.

\section{Introduction}
Cell motility drives many biological processes, including morphogenesis and wound healing, and its disregulation is implicated in diseases like cancer. In the past, cell motility research has often focused on the behavior of single cells, such as fibroblasts, in different environments~\cite{Petrie2015,Tschumperlin2013}.


Recently, there has been a growing understanding that in dense tissues cell motion can be driven by collective effects, i.e. by cell-cell interactions instead of cell-autonomous properties. One particular area of focus has been "cellular jamming", a term that researchers in biology and related fields have adopted to describe the collective arrest of cell motion in dense tissues~\cite{Haeger2014,Sadati2013,Park2015,Kim2020,Angelini2011}. One reason the concept is useful is because it suggests new, non-cell autonomous mechanisms can alter cell motion, potentially identifying new targets for therapies for disease. For example, recent work emphasizes that non-cell autonomous processes, such as cell-cell adhesion~\cite{Mongera2018,Kim2020,Petridou2021} and stress fluctuations driven by nearby cell division~\cite{Oswald2017,Saadaoui2018, Czajkowski2019}, impact cell motility and structural rearrangements in dense tissues. 

While it is exciting that a widening group of researchers are studying the collective arrest of cell motion, a challenge is that the term "cell jamming" is being used as a broad umbrella description of such processes. Since there are multiple distinct mechanisms that can drive collective cell arrest, and the term "cell jamming" has been used to describe all of them, it often remains unclear which mechanisms are actually operating in a given process. Therefore, the focus of this review is to describe several distinct mechanisms for collective cell arrest, and highlight ideas for how one might confirm a given mechanism is operating in a given situation.


To build intuition about mechanisms for collective arrest in cells, we turn to the physical sciences, where the collective arrest of particle or molecular motion is termed "solidification" or "rigidification". In introductory physical science classes, one learns that a material can be solidified by cooling -- i.e. reducing fluctuations induced by temperature -- or by increasing pressure -- i.e. packing the particles, atoms, or molecules closer together. When a material is cooled into a solid while remaining disordered, it undergoes a glass transition. And in the physical sciences and engineering, jamming is a technical term reserved to describe the onset of solidification at zero temperature, driven specifically by changes to pressure or density. Researchers have developed a "jamming phase diagram" to unite various mechanisms that are responsible for solidification~\cite{Trappe2001,Andrea1998}. 

In this review, we take our cue from the physical sciences and adapt recent results from the literature to construct a jamming phase diagram for cell collectives, focused on three mechanisms: crowding, active fluctuations, and tension-driven rigidity. We are not the first to conjecture such a diagram; several other phase diagrams have been proposed previously~\cite{Park2015, Bi2016}.  Nevertheless, recent work over the past two years has generated new explicit predictions for the onset of cell arrest and provided experimental evidence for its validity. Perhaps more importantly, it has become clear that in real tissues, multiple mechanisms that could drive cell arrest are often operating at the same time in subtle ways. Therefore, it is not sufficient for scientists to measure a single quantity, such as cell number density, and claim that changes in that quantity are driving cell arrest.  Instead, as highlighted in the work below, it is important to quantify multiple observables quantitatively in space and time to confirm the dominant mechanisms driving cell arrest.



\begin{figure}[]
\includegraphics[width = 0.8\linewidth]{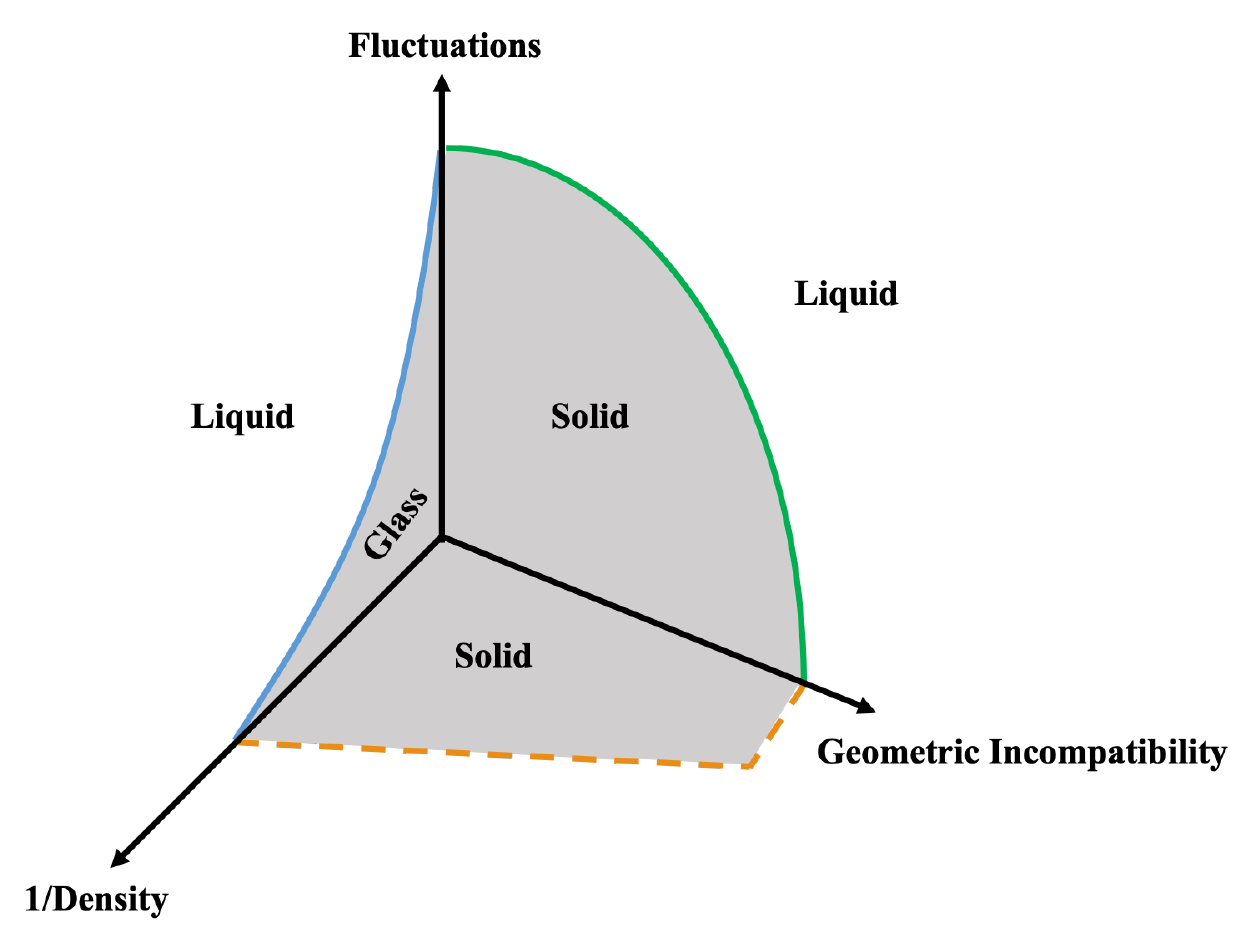}
\caption{A speculative cell jamming phase diagram extrapolated from recent results in literature. The green line represents the tension-driven rigidity transition seen in confluent tissues due to a competition between fluctuations and cell shape induced geometric frustration~\cite{Bi2016}. The blue line represents the glass transition seen in harmonic spheres~\cite{Berthier2011a}, which models the behavior seen by nonconfluent rounded cells at low adhesions. The orange line represents the shear instability seen in partially confluent tissues at finite temperature as density and adhesion are tuned. At low adhesion, the tissue becomes solid-like as density increases, reminiscent of crowding. At high adhesion, there is a density-independent transition similar to what is seen in completely confluent tissues~\cite{Kim2020}.}
\label{fig:figure1}
\end{figure}

\section{Physical mechanisms for cell arrest}
In this brief review, we will follow the existing literature and use the term "cell jamming" to refer to a collective arrest of cell motion.  In jammed or solid-like tissues, cells do not change neighbors, the cell-scale structure does not remodel, and the tissue resists changes to its shape.  In unjammed or fluid-like tissues, cells do change neighbors, and the tissue flows and remodels in response to fluctuations or internally or externally applied forces.

We next discuss in some detail three mechanisms that can drive jamming or unjamming: crowding, tension-driven rigidity, and fluctuations.

\begin{figure}[ht]
\includegraphics[width = 0.75\linewidth]{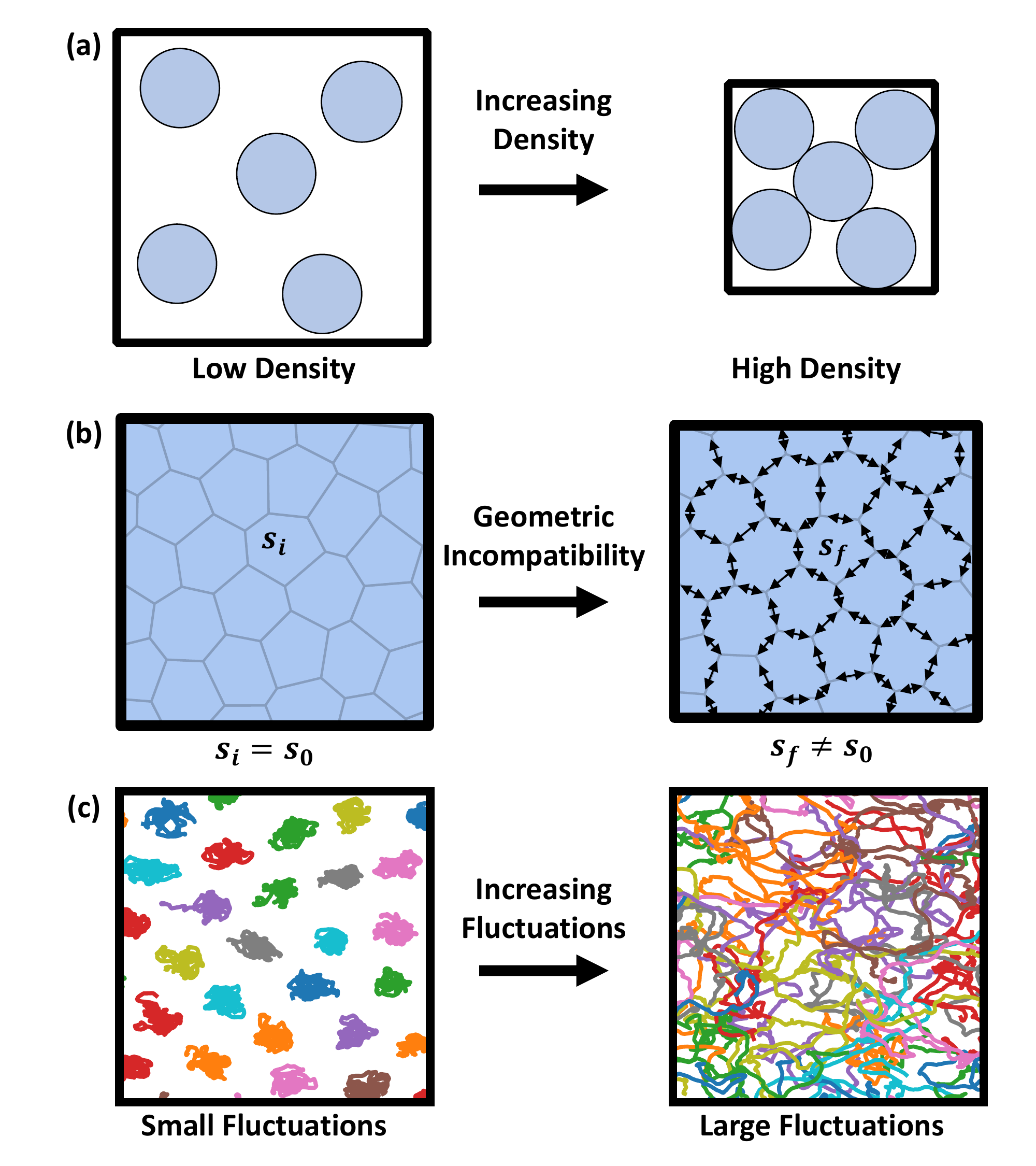}
\caption{General mechanisms for cell arrest. (a) Crowding: Compression of cells inside a box increases the number of contacts, and therefore constraints, on each cell which causes the cells to jam. (b) Tension-driven rigidity: Cells in a monolayer are fluid-like when their current cell shape, $s_i$, is the same as their preferred shape, $s_0$. Then by altering the preferred cell shape, the cells become geometrically frustrated and the tissue becomes rigid. (c) Fluctuations: Trajectories of cells in a monolayer illustrate the caging effect of cells at low temperature. As the temperature is increased, the cells have the energy to escape their cage and rearrange.}
\label{fig:figure2}
\end{figure}

\subsection{Crowding} 
Crowding is a mechanism for cell jamming that is directly related to particle jamming in the physical sciences. Crowding occurs when the fraction of available space taken up by particles, molecules, or cells becomes sufficiently high that the entire system becomes rigid. It is easiest to understand the math behind this mechanism in the simplest case of squishy spheres at zero temperature.  Each sphere can move left or right, forward or backward in two dimensions (and also up and down in three dimensions). Therefore, the number of degrees of freedom in the system is the number of particles $N_p$ times the number of dimensions $d$: $ N_{DOF}= N_p d$. 

When things get crowded, particles must start to contact other particles, and each contact adds another constraint to the system, shared between the two particles. So if each particle has on average $z$ contacts, the number of constraints in the system is $N_c = N_p z/2$. In line with intuition, the system rigidifies precisely when the number of constraints equals the number of degrees of freedom: $N_c = N_{DOF}$, which can also be written in terms of the average number of contacts as $z_{critical} = 2 d$.  Therefore, when the system is less crowded and the number of contacts is less than this critical value, the system is floppy and particles can change neighbors. When the system is sufficiently crowded so that the number of contacts is above that critical value, the system is rigid.

Of course, real materials deviate from this idealized case. Making the spheres slightly adhesive alters the nature of the constraints and changes the rigidity transition~\cite{Trappe2001}. In the presence of finite fluctuations, such as a non-zero temperature, the particles can still be "caged" by their neighbors, and that state is called a glass. In the presence of applied forces, a system rigidified by crowding can also begin to yield and flow. Understanding the precise nature of these glasses and yielding transitions is still a highly active field~\cite{Berthier2011,Liu2010}.


Given that many cell types round up in cell culture medium and respond much like sticky "active bubbles"~\cite{Evans1989}, it is obvious how a crowding mechanism could generate rigidity in a tissue. Roughly spherical cells could adhere, divide, or be compressed until the density of cells increases past the critical threshold and they can no longer move past each other. Modeling cells as sticky spheres generate a jamming phase diagram that depends on density, adhesion, and applied stress~\cite{Trappe2001} which was the inspiration for a conjectured phase diagram for cell arrest~\cite{Sadati2013}. 

A very clear example of this "sticky-sphere" transition was recently discovered in the zebrafish blastoderm~\cite{Petridou2021}, a tissue where the viscosity drops by more than an order of magnitude in a few minutes, as the tissue transitions from a solid-like to a fluid-like state. During this transition, it is observed that the cell packing fraction decreases slightly, while the viscosity drops precipitously. This is precisely what is seen in particulate jamming, where a small change in packing fraction drives a significant change in the contact network or connectivity. In the new work, the authors meticulously reconstruct cell connectivity networks to demonstrate the same effect as in jammed particles, and then perturb E-cad expression and demonstrate that the resulting change in network connectivity completely explains the resulting change in tissue viscosity.

The picture becomes more complicated in heterogeneous systems. For example, recent work demonstrates that a fluid-to-solid transition occurs as a function of position along the body axis in zebrafish embryo~\cite{Mongera2018}. Fluid-like cells in the mesodermal progenitor zone (MPZ) differentiate and are incorporated into the presomitic mesoderm (PSM). The MPZ has larger extracellular spaces and active fluctuations than the PSM leading to the PSM having a higher cell density. The MPZ tissue is fluid-like with a lower tissue viscosity and lower yield stress than the PSM. Together this is suggestive that the embryo experiences a jamming transition due to crowding along the anterior-posterior axis.

\subsection{Tension-driven rigidity} 
While the standard explanation for rigidity in materials is crowding, a new type of rigidity transition has recently been discovered in biological systems, such as confluent tissues and biopolymer networks, and also in so-called "mechanical metamaterials" like origami~\cite{Chen2018}. It is deeply connected to the older idea of tensegrity structures as models for cells and tissues~\cite{Ingber2003}.

In all of these systems, the contact network (sometimes called the network topology) remains constant, in direct contrast to jamming scenarios. Instead, a continuous parameter can be tuned so that the system crosses a rigidity transition. The underlying mechanism has been termed \emph{geometric incompatibility}~\cite{Moshe2018,Merkel2019} and it is similar to how a guitar string becomes rigid once it is stretched beyond the initial length of the string. We will refer to this type of transition as "tension-driven rigidity".



In biopolymer networks, the tuning parameter is the applied strain (amount of external deformation). For small strains, the network is floppy, and at a critical strain the stiffness of the network changes by several orders of magnitude. This behavior is predicted by simple models and observed in experiments~\cite{Storm2005,Koenderink2009,Rens2016}.


Similarly, this rigidity transition can also be seen in models for confluent tissues. In vertex or Voronoi models for such systems, each cell has a characteristic volume (or area in 2D cross-sections of monolayers). In addition, each cell has a preferred surface area (or perimeter in 2D), which is generated by cell-cell adhesion due to cadherins and other adhesive molecules, a surface-area minimizing cortical shell of actin, myosin and other cytoskeletal components, and non-linear effects such as contractile rings or saturation of molecules at adhesive contacts. In these models, the dimensionless preferred cell shape, which in 2D is just the ratio of the preferred perimeter to the square root of the preferred area, continuously tunes the model across the transition~\cite{Farhadifar2009, Bi2015}, in agreement with what is seen in experiments~\cite{Park2015}.  Recent work has demonstrated that these models quantitatively predict, with no fit parameters, cell rearrangement rates in body axis elongation in the fruit fly~\cite{Yan2019, Wang2020}, if one also takes into account cell alignment and disorder in the packing. Similar effects are predicted in 3D~\cite{Merkel2018}.


\subsection{Interpolating between crowding and tension-driven rigidity:}
While crowding is often studied in particle-based models where the interaction depends on how much the particles overlap, and tension-driven rigidity has been studied on vertex networks that fill all of space, in real biological systems we are often interested in tissues that are nearly confluent, with small gaps between cells. Do such tissues rigidify due to crowding, tension-driven rigidity, both, or something else?

One recent manuscript~\cite{Huebner2020} showed that in nearly confluent mesenchymal tissues in Xenopus development, cells exhibited features of both types of systems -- they could actively tug past other cells (more like particle-based crowding models) and they could also shorten and extend edges to change neighbors more like vertex models. 


To explain such observations, a new set of partially confluent models have recently been developed. One version begins with a particle-based model so that cells can fully break apart, but where overlap between cells can create an interfacial edge~\cite{Teomy2018}. Due to a competition between two-cell interactions, three-cell interactions, and geometric constraints, this model exhibits several different tissue phases, including a gas phase where cells behave as repulsive spheres and confluent phases that shares features with those found in vertex models. Another set of simulations investigates cells modeled as deformable rings that can change shape up to the confluent limit~\cite{Boromand}. In this model, the tissue always behaves like an elastic solid with invaginations occurring after confluence. As of yet, neither model has been directly compared with experiments.


Most recently, an exciting new study developed an active foam model, which specifies a foam-like interfacial tension on each edge of a cell, and allows gaps to open up spontaneously if they are energetically favored.
This model replicates both the crowding transition seen in passive foams with increasing packing fraction and some aspects of the tension-driven rigidity transitions seen in vertex models~\cite{Kim2020}.  However, the model includes only linear interfacial tensions and does not include nonlinear effects that stabilize the fluid phase of vertex models, so it will be interesting to see how nonlinear effects might alter predictions.

\subsection{Role of fluctuations:}
Finally, we return to the familiar idea of a material becoming solid or glassy as the temperature is decreased. Temperature fluctuations provide an energy source that allow a particle to explore its local environment, and a particle inside a dense system requires energy to escape the constraints imposed by its neighbors -- i.e. break out of its cage.  As the temperature is decreased, it becomes less likely that a particle will have the energy required to do so. 

Similarly, in self-propelled particle models for tissues, where a particle generates its own forces and momentum, the system can also undergo a glass transition~\cite{Henkes2011} controlled by the packing fraction (as in crowding) as well as the persistence time (how long a particle moves in a straight line before changing direction) and the magnitude of the self-propulsion forces. Detailed work has shown that the glass transition is different in thermal systems compared to self-propelled ones~\cite{Berthier2011}.


Fluctuations also play a similar role in confluent models. In self-propelled confluent models, as the magnitude of the propulsion force or the persistence time decreases, the system becomes more solid-like~\cite{Bi2016}. Glassy behavior is observed in the low-temperature fluid phase of the vertex model, although a detailed analysis reveals that some features are interestingly different from what is observed in particulate systems~\cite{Sussman2018}. Another possible source of fluctuations is fluctuating tensions along cell-cell interfaces, which have been shown to be coupled with the cycle of cell division~\cite{Jones2018}. Recent simulations~\cite{Krajnc2020, Yamamoto2020} and experiments~\cite{Devany2019} suggest that a confluent tissue can solidify as the magnitude of tension fluctuations in the tissue decrease, although there is evidence that the persistence of tension affects tissue fluidity in a complicated way~\cite{Yamamoto2020}. 

Finally, recent experiments on pharmacologically perturbed epithelial cells highlight the possibility that cell arrest is not associated with an underlying rigidity transition at all. In the perturbed cell types, it appears that the relevant fluctuations become so small that cells do not change neighbors and remodel even though the cellular structure is floppy~\cite{Devany2019}. In other words, the cells could move easily, but they do not because there are no fluctuations to drive them, similar to an unjammed particulate system at absolute zero temperature. This highlights that even in dense tissues, the absence of cell motion is not automatically a proxy for solidity or rigidity.


\section{Biological drivers of jamming and cell arrest}
In the previous section we focused on the physical mechanisms that govern cell arrest, but clearly those physical mechanisms are driven by specific cell-scale and molecular-scale features.  In addition, individual cells can sense and respond to their environments in mechanosensitive feedback loops.  In this section, we highlight how specific cell-scale biological features such as expression levels and signaling pathways can be used to tune cell arrest in the scenarios discussed above. 


\subsection{Cell-cell interactions}
  Multicellular organisms have evolved a large number of redundant cell-cell interactions that act to preserve tissue cohesion and allow the robust formation of tissue-scale structures. Many of these signaling cascades begin with cell-cell adhesion molecules, such as cadherins.  We first focus on homotypic interactions between two cells of the same type. In non-confluent tissues, evidence is emerging that cell-cell adhesion performs a role very similar to that expected for adhesion in sticky spheres~\cite{Trappe2001}; decreasing adhesion generates larger intracellular spaces and fewer cell-cell contacts, which increases the fluidity of the tissue via a decrease in crowding~\cite{Mongera2018,Kim2020,Petridou2021}.  
  
  In confluent tissues, the role of cell-cell adhesion is much more subtle and likely cell-type specific. In such tissues, the tuning parameter for tension-driven rigidity is cell shape, and evidence is emerging that cadherin expression levels can change the cell shape in context-dependent or non-monotonic ways.  For example, cell doublet experiments suggest that, consistent with intuition, increasing cadherin expression increases the surface area of cell-cell contacts~\cite{Yamada2007}. However, new experiments in confluent monolayers show that knockdown of E-cadherins in keratinocytes results in an increase in cell shape index compared to wildtype~\cite{Sahu2020}, which indicates that cells with lower E-cadherin expression prefer \emph{more} surface area of cell-cell contact. In some ways, these subtle behaviors are not surprising, as it is well-known that cadherin signaling significantly changes the mechanics of the cortical cytoskeleton~\cite{Maitre2013,Amack2013} so that increases in adhesion are often balanced by changes to cortical tension that have an opposite effect.
  
  
  A related phenomenon is heterotypic cell-cell interactions between two different cell types. Mixtures of two cell types are often observed to sort, and the mechanisms driving such sorting should be deeply related to crowding-based vs. tension-driven rigidity transitions discussed in the previous section. For example, the differential adhesion hypothesis (DAH) is based on the assumption that cells act like sticky spheres, with an interfacial tension directly proportional to differences in cell-cell adhesion, so sorting occurs when cells rearrange to minimize that interfacial tension~\cite{M.S.Steinberg1963}. 
  
  In confluent models, on the other hand, adhesion-based changes in cell shape are not sufficient to drive macroscopic sorting~\cite{Sahu2020}. Instead, confluent models require a specific heterotypic response -- an explicit change to interfacial tension along heterotypic cell-cell contacts -- in order to generate macroscopic sorting~\cite{Canty2017,Sussman2018a}. Heterotypic interfacial tension causes complete and rapid demixing between cell types~\cite{Sahu2020} and there is a discontinuous restoring force for perturbations of the boundary~\cite{Sussman2018a}. For tissues near the fluid-solid transition, the final cell and interface shapes may be set by a competition between the interfacial forces and the shape-based forces governing tension-driven rigidity~\cite{Sahu2021}.




  Another class of cell-cell interactions -- particularly important in morphogenesis -- are pathways such as planar cell polarity that localize adhesion molecules and motor proteins along interfaces with a specific orientation, generating large-scale anisotropic forces~\cite{Zallen2004,Bertet2004}.  It has recently been shown that any anisotropic forces, including those generated by external stretching or pulling from nearby tissues~\cite{Duda2019}, alter tension-driven rigidity. Specifically, while previous work focused largely on isotropic systems where cells are not aligned, anisotropic forces generically lead to cell alignment, where the long axis of cells point in the same direction. By carefully measuring cell shape, cell-cell alignment, and the disorder~\cite{Yan2019}, one can extend theories of tension-driven rigidity to predict rates of cell rearrangement with no fit parameters. Remarkably, these predictions with no fit parameters quantitatively match experimental data for body axis elongation in Drosophila~\cite{Wang2020}.

  
  In addition to changing the overall magnitude of tension on a cell-cell interface, cell-cell interactions can also induce fluctuations in intercellular tension, as well as feedback loops to regularize such fluctuations.
  For example, pulses of non-muscle myosin II have been shown to cause permanent junctional remodeling that can drive shape changes and increased cell rearrangements during convergent extension. A recent model captures this behavior by allowing for junctions to undergo permanent tension remodeling after surpassing the critical strain threshold. However, to avoid permanent junctional shortening, there is continuous strain relaxation which allows the system to slowly lose deformation memory. Together this allows for large-scale irreversible deformations during convergent extension~\cite{Staddon2019}. A different model allows for cytoskeletal remodeling through active recruitment of myosin depending on the internal strain rate of its associated actin filament.  In this case, myosin pulsation causes deformations in cell shape, which in turn stimulate myosin recruitment, which then stabilizes the deformation~\cite{Noll2017}. Additionally, the adhesion molecule Sidekick (sdk) is shown to localize at tricellular adherens junctions (tAJs) and disruptions to sdk cause abnormal cell shape changes and a decrease in rearrangements contributing to convergent extension. One hypothesis for this behavior is that the sdk adhesion molecule is involved with the transition from shortening to elongation which occurs at tricellular vertices during intercalation. To capture this, the authors developed a vertex model where higher-fold rosettes structures were stabilized and long-lived, and the simulations generated shape changes and intercalation rates that were quite similar to the sdk mutants. This suggests that sdk may exert feedback control of tension fluctuations at junctions, and disruptions in sdk may significantly delay or halt cell rearrangements~\cite{Finegan2019}.
  
   A related observation is that many cell types exhibit active forces, such as cell motile forces or tension along stress fibers, that are polarized along a specific direction. Cell shape alignment and other types of cell-cell signaling can drive these polarizations to align in the same direction.  Such alignment interactions can lead to large-scale collective behavior~\cite{Vicsek2012} and can alter cell jamming in dense tissues. For example, in crowding models, cells become more aligned as the packing fraction increases preceding the onset of rigidity~\cite{Henkes2011}. In confluent models, increasing alignment in cell polarity drives the tissue towards a "solid-flocking" state~\cite{Giavazzi2018}, very similar to the behavior seen in epithelial monolayers with upregulated RAB5A~\cite{Malinverno2017}. This observation highlights that not all solid-like states have arrested motion: the "solid-flocking" state corresponds to a group of cells that is internally rigid so that the cells do not change neighbors, and yet the collective is still moving together in the same direction as a unit.  Therefore, one needs to examine the relative displacements between cells rather than the absolute displacement of cells to determine tissue fluidity.
  


\subsection{Cell-substrate interactions}

Cell-substrate interactions play a key role in cell jamming, as many cell types exert forces on the substrate in order to locomote, via integrin-based traction forces~\cite{Huttenlocher2011}. Especially in epithelial monolayers, these active propulsion forces generate fluctuations that can drive fluidization of the tissue. In less confluent systems, cell-cell contacts may decrease this propensity through contact inhibition of locomotion~\cite{Zimmermann2016}.

The mechanical and biochemical properties of the substrate also provide cues that can alter jamming and cell arrest.  For example, cells become more stiff on stiff substrates~\cite{Gupta2015,Gupta2019,Prager-Khoutorsky2011}, which would be expected to lead to higher energy barriers in particle-based crowding models and generate changes to preferred cell shape in confluent models. Simultaneously, cells tend to spread more on stiffer substrates~\cite{Guo2017}, which would decrease cell number density in particle-based models and decrease cell shape index in confluent models, in the absence of other feedbacks. Adhesion matters too -- increasing cell-substrate adhesion enhances cell spreading and drives high locomotion~\cite{Cao2019}. A very valuable direction for future research would be to characterize how cell-cell adhesion and cell-substrate adhesion co-regulate one another and thereby alter cell arrest.



\subsection{Cell Division}
Historically, cell division has been highlighted as a mechanism for increasing cell densities in tissues. Specifically, in particle-based models with fixed boundaries where cells grow before dividing, cell divisions increase crowding which in turn leads to cell arrest.  However, there are other biophysical mechanisms triggered during cell division that can affect tissue rigidity.

First, the act of cell division necessarily creates active stress fluctuations that can fluidize the tissue~\cite{Oswald2017,Saadaoui2018, Czajkowski2019}. Second, tension fluctuations are often introduced, for example, by asymmetric division where daughter cells have different mechanics, as seen in mouse blastocysts~\cite{Maitre2016}, on in symmetric divisions which result in lower tension between mitotic cells and their neighbors compared to other interfaces~\cite{Miroshnikova2018}. Lastly, there is emerging evidence that stereotyped changes to global cortical tension occur as a function of cell cycle~\cite{Jones2018}. Recent work has demonstrated that this last mechanism is likely the dominant source of fluctuations and tissue fluidization in MDCK monolayers~\cite{Devany2019}.





\subsection{Cell differentiation}
Cell differentiation can also drive changes to tissue rigidity. As higlighted previous previously,  the differentiation of cells from the MPZ to PSM in zebrafish embryos results in smaller gaps between cells and jamming due to crowing~\cite{Mongera2018}. This fluid-solid transition along the body axis guides the morphogenetic flows that shape the embryo~\cite{Banavar2020}. 

Similarly, in cancers, the epithelial to mesenchymal transition (EMT) causes epithelial cells to become less confluent and more migratory~\cite{Yang2020,Tripathi2020}. In mixtures of these two cell types, which occur when only a fraction of the cells have transitioned to a mesenchymal type, it is observed that increasing the fraction of mesenchymal cells results in an increase in motility and cell shape of the epithelial cells. This frustrates jamming in the tissue~\cite{GamboaCastro2016}.


\section{Concluding Remarks}
The goal of this review is to reduce some of the ambiguity around the increasingly common term "cell jamming". We highlight that jamming and the collective arrest of cell motion can be driven by multiple physical mechanisms, and emphasize the role of three specific mechanisms: crowding, tension-driven rigidity, and a reduction in fluctuations. We develop a "cellular jamming" phase diagram along these three axes extrapolating from recent results in the literature. Careful quantitative measurements are required to distinguish which mechanism is dominant, and recent work has begun to carefully test where in this phase space different tissues operate.

In addition, we attempt to connect specific cell-scale features, such as the expression levels of cadherins or frequency of cell divisions, to the physical mechanisms that appear in the cell jamming phase diagram. However, given the scope, complexity, and cross-talk between these cell biology processes, much work remains to be done to understand how specific molecular mechanisms are connected to the physics of tissue rigidity. Moreover, cells can sense the rigidity of the surrounding tissue and alter molecular-scale properties in response. Using cell jamming as a lens to understand how such feedback loops guide morphogenetic processes and how the disregulation of such feedback loops drive disease will be an exciting direction for future research.


\section{Conflict of interest statement}
The authors declare no conflicts of interest.

\section{Funding Sources}
This work was supported by the Simons Foundation grants $\#454947$ and $\#446222$ and the National Science Foundation NSF-DMR-1951921.

\section{Annotated References}
Papers of particular interest, published within the period of review, have been highlighted as

$\ast$ of special interest

$\ast \ast$ of outstanding interest

\begin{itemize}
    \item $\ast \ast$ Devany 2019~\cite{Devany2019}: Cell division Rate Controls Cell Shape Remodeling in Epithelia 
\end{itemize}

This paper observes the collective motion and rheology of MDCK monolayers after reaching confluence. On different ECM stiffnesses, the number density of the tissues where cell motion reaches its minimum varies significantly. This highlights that crowding alone isn't always enough to predict jamming behavior.  Also, researchers observed that disruptions to the cell cycle cause collective cell arrest. Together with their measurements of cell-cell contact angles, this suggests that cell-cycle dependent fluctuations in interfacial tension are responsible for tissue remodeling.

\begin{itemize}
    \item $\ast \ast$ Kim 2020~\cite{Kim2020}: Embryonic Tissues as Active Foams
\end{itemize}

This paper describes an active foam model for near-confluent tissues where gaps are allowed to open spontaneously. The model replicates both the crowding transition seen in passive foams with increasing packing fraction and some aspects of the tension-driven rigidity transitions seen in vertex models. Their results suggest that tension fluctuations rather than adhesion or cortical tension may control tissue rheology in some embryonic tissues.

\begin{itemize}
  \item $\ast$ Mongera 2018~\cite{Mongera2018}: A fluid-to-solid jamming transition underlies vertebrate body axis elongation. 
\end{itemize}

This paper highlights the fluid-solid transition along the body axis in the zebrafish embryo. The fluid-like cells in the mesodermal progenitor zone have larger extracellular spaces and fluctuations than the solid-like presomitic mesoderm. This is suggestive that the embryo undergoes the jamming transition due to crowding along this anterior-posterior axis. Together with~\cite{Banavar2020}, the results suggest that the fluid-solid transition may be responsible for guiding morphogenetic flows necessary for embryonic morphogenesis.

\begin{itemize}
    \item $\ast \ast$ Petridou 2021~\cite{Petridou2021}: Rigidity percolation uncovers the structural basis of embryonic tissue phase transitions
\end{itemize}

This paper investigates the rigidity transition seen in the zebrafish blastoderm. The large drop in the tissue's viscosity is explained by changes to the cell contact network, which are strikingly similar to those predicted by theories of rigidity percolation that explain jamming in soft spheres. The results indicated that when E-cadherin expression is reduced, the resulting decrease in cell-cell adhesion reduced connectivity in the contact network and increases tissue fluidity.  Together this indicates that in this system cellular jamming is caused by increased crowding driven by increases in cell adhesion. Interestingly, it's also observed that synchronous cell divisions are required for homogeneously distributed contact loss throughout the tissue, which is necessary to produce uniform changes in viscosity.

\begin{itemize}
    \item $\ast$ Staddon 2019~\cite{Staddon2019}: Mechanosensitive Junction Remodeling Promotes Robust Epithelial Morphogenesis
\end{itemize}

This paper demonstrates that tissue-scale shape changes during convergent extension can be driven by frequency-dependent actomyosin pulsation. Experimental observations indicate that epithelial junctions behave elastically under low contractile stress but undergo reversible deformations under high contractile stress. The authors propose an extension to the vertex model in which epithelial junctions remodel after a critical strain is surpassed and impose a continuous strain relaxation. Together these additions capture the tissue remodeling seen in experiments. 

\begin{itemize}
    \item $\ast$ Wang 2020~\cite{Wang2020}: Anisotropy links cell shapes to tissue flow during convergent extension 
\end{itemize}

This paper examines cell anisotropy induced during convergent extension of the Drosophila germband epithelium. It's observed that tension-driven rigidity predictions from cell shape alone do not predict the rearrangement rates inside the tissue. A combination of cell shape and cell anisotropy is shown to robustly predict cell rearrangement rates and tissue fluidity with no fit parameters body axis elongation in Drosophila.

\begin{itemize}
    \item $\ast$ Yan 2019~\cite{Yan2019}: Multicellular Rosettes Drive Fluid-solid Transition in Epithelial Tissues 
\end{itemize}

This work finds that the rigidity of epithelial tissue increases as the number of multicellular rosettes increases. It predicts a scaling relation describing how global tissue tension grows as higher-order vertices are created and the tension of cell edges adjacent to rosettes increases. Their predictions of cell topology and edge tensions closely match with experimental observations in Drosophila embryos. They also develop a generalized Maxwell constraint counting approach to explain the rigidity transition seen in confluent tissue.

\bibliographystyle{plain}
\bibliography{CellJammingReview}

\end{document}